\documentclass[a4paper, superscriptaddress, twocolumn, prl, showkeys]{revtex4-1}

\usepackage[colorlinks=true,citecolor=blue,linkcolor=blue,pdfhighlight =/O]{hyperref}
\usepackage{graphicx}
\usepackage{amsmath}
\usepackage{mathtools}
\usepackage{epstopdf}
\usepackage{color}
\usepackage{soul}
\usepackage{url}
\usepackage{natbib}

\hypersetup{urlcolor=blue}

\begin{document}

\title{Probing Hybridization of a Single Energy Level Coupled to Superconducting Leads}
\author{D. M. T. van Zanten}

\author{H. Courtois}

\author{C. B. Winkelmann}

\email{clemens.winkelmann@neel.cnrs.fr}
\affiliation{Univ. Grenoble Alpes, Inst. NEEL, F-38000 Grenoble, France}
\affiliation{CNRS, Inst. NEEL, F-38000 Grenoble, France}

\date{\today}

\begin{abstract}

Electron transport through a quantum dot coupled to superconducting leads shows a sharp conductance onset when a quantum dot orbital level crosses the superconducting coherence peak of one lead. We study superconducting single electron transistors in the weak coupling limit by connecting individual gold nanoparticles with aluminum  junctions formed by electromigration. 
We show that the transport features close to the conductance onset threshold can be accurately described by the quantum dot levels' hybridization with the leads, which is strongly enhanced by the divergent density of states at the superconducting gap edge. This highlights the importance of electron cotunneling effects in spectroscopies with superconducting probes.

\end{abstract}


\maketitle

The transport properties of a tunnel junction to a single quantum dot (QD) are a sensitive probe of the its spectral properties. As the bias voltage across the junction is ramped up, new quantized transport channels - associated to the QD energy levels - become accessible, leading each to a stepwise increase in the current \cite{Beenakker91,Averin91,Ralph95,Kouwenhoven97}. The electronic tunneling process associated to each channel can be elastic - reflecting thereby the QD energy spectrum - or inelastic,  in which case additional bosonic excitations are involved \cite{Park00, DeFranceschi01,Heinrich13}. When normal conductive leads are used, thermal broadening limits spectroscopic resolution to  about $3.5\, k_BT$ in weakly coupled QD junctions. The use of superconducting (S) leads strongly modifies the picture. As soon as the thermal energy $k_BT$ is a few times smaller than the superconducting gap $\Delta$, thermal quasiparticles in the leads have a vanishingly small density. Thermal smearing is thereby avoided and the spectroscopic resolution can be much improved \cite{Pan98,Rodrigo04,Grove09,Pillet10,Franke11,Heinrich13,Nadj14}. 
Ultimately, another limitation to spectroscopic resolution arises from the tunnel coupling induced hybridization of the isolated quantum level at study. With normal contacts, this manifests as a lorentzian broadening of the spectral features, set by the tunnel coupling energy scale $\hbar\gamma$. 
Nevertheless, this description does not hold in the case of superconducting leads \cite{Yeyati97,Kang98}.

\begin{figure}
  	\includegraphics[width=1.0\columnwidth]{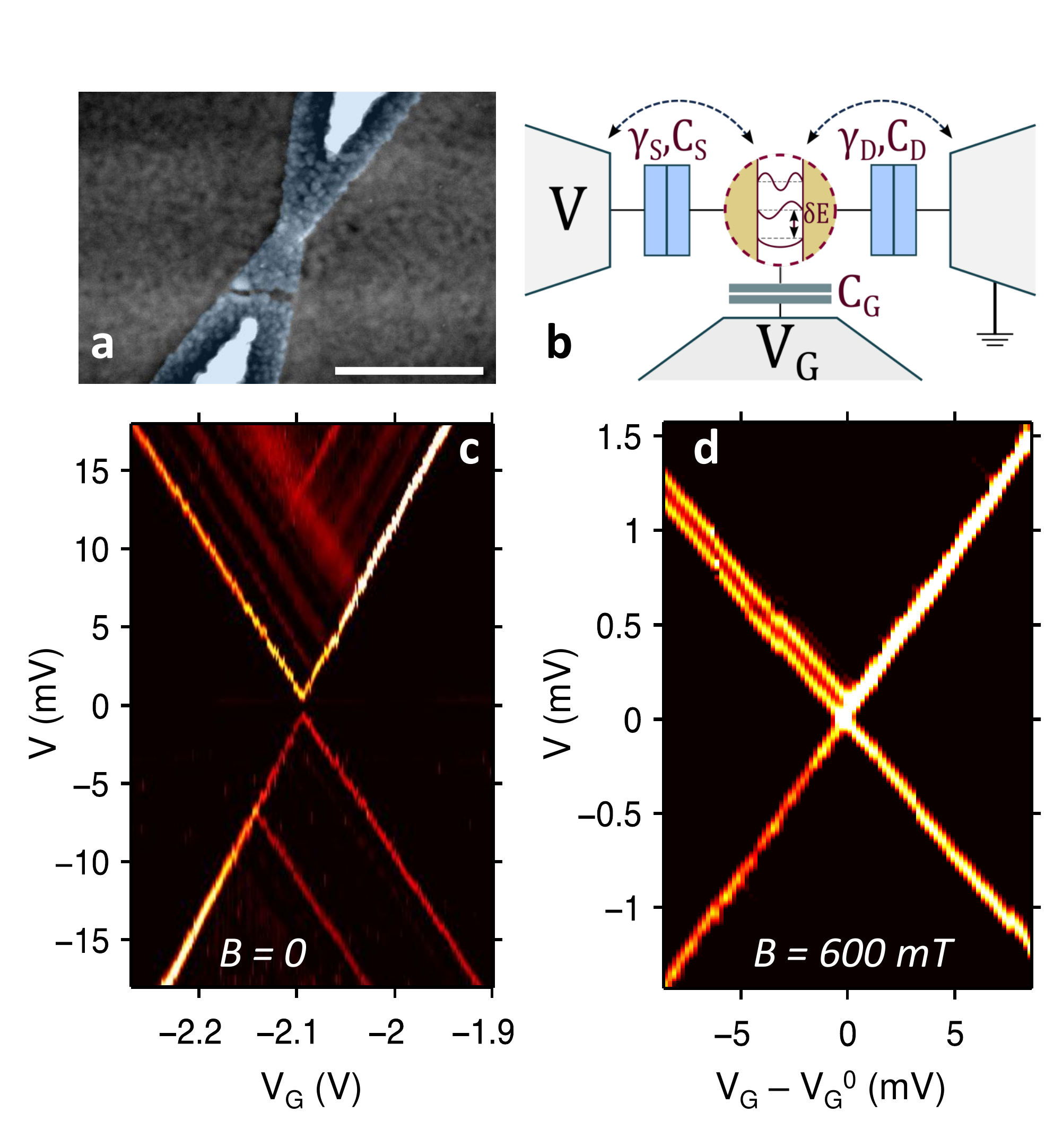}
	\caption{(a) Scanning electron micrograph of bare Al constriction after electromigration. The scale bar is 250 nm (no nanoparticles present here). (b) Schematic of the S-QD-S device, introducing the different capacitances and tunnel couplings. (c) $\partial I / \partial V$ differential conductance map of the asymmetric device A, displaying its sole experimentally accessible degeneracy point. The conductance ranges from about 0 (black) to 1.6 $\mu$S (bright) ($T=100$ mK, zero magnetic field). A gap at the degeneracy point is visible. (d) Zoom on the degeneracy point at device A at a field of 600 mT. The gap has disappeared and the conductance onset is Zeeman split at one Coulomb diamond edge.}
	\label{device}
\end{figure}

 In this Letter, we report transport measurements in S-QD-S junctions in the weak coupling limit, in a regime where a unique separation of all relevant energy scales allows for quantitative comparison to theory.
We experimentally show that the large DOS at the superconducting gap edge of the leads strongly enhances hybridization of the single quantum level it is tunnel coupled to.
Hybridization of the QD spectral function is found to be the dominating limitation to spectroscopic resolution, despite the weak coupling.

The device fabrication process relies on controlled electromigration of an on-chip all-metallic device presenting a constriction \cite{Park99}. This technique was successfully applied for connecting single molecules in numerous experiments \cite{Park00, Park02,Liang02, Osorio08,Nitzan03,Yu04}.  Using aluminum constrictions, electromigration has been used for forming superconducting single molecule transistors \cite{Winkelmann09}. In that experiment, the use of a narrow section of proximity superconducting gold to contact the molecular QD yielded rather high tunnel couplings, at the cost however of losing a hard superconducting gap in the leads. Alternatively, pure aluminum leads provide more weakly coupled devices, with BCS type superconducting contacts.
We pattern aluminum constrictions $\sim$ 100 nm wide and 15 nm thick on top of a locally defined backgate. Optimal gate coupling along with galvanic isolation up to gate voltages $\sim8$ V is provided by a 10 nm thick atomic layer deposition of alumina on top the backgate electrode. Gold nanoparticles of 5 nm diameter \cite{nanocomposix} are dry-casted onto the sample from a toluene suspension shortly before introducing the sample into the cryostat. We perform electromigration at 4.2 K in the cryogenic vacuum of a home-made inverted dilution refrigerator. A bias voltage is slowly ramped up ($\sim$ 30 mV/s) while monitoring the constriction conductance. A real-time controller sets the bias to zero within microseconds as soon as the constriction conductance falls below 100 $\mu$S. In spite of the high access series resistance of 170 $\Omega$ per wire, which is due long lossy coaxial lines at base temperature on each lead for filtering, electromigration typically sets on at about 1 V. Electromigration gaps are difficult to see in the Scanning Electron Micrsocope, especially in the presence of the spin-coated nanoparticle layer. Figure \ref{device}a shows the image of a large gap with virtually infinite tunnel resistance for the purpose of illustration.

 After electromigration, most junctions display tunnel behavior with resistances in the upper M$\Omega$ range. Samples showing stable gate-dependent conductance features are further investigated {\em in situ} at dilution temperatures. All measurements reported here were performed at an electronic base temperature $T\approx 100$ mK.  The current-voltage $I(V)$ characteristics are measured in a heavily filtered DC transport environment; differential conductance $\partial I / \partial V$ maps at different gate voltages $V_G$ are obtained by numerical differentiation.

\begin{figure}
	\includegraphics[width=0.9\columnwidth]{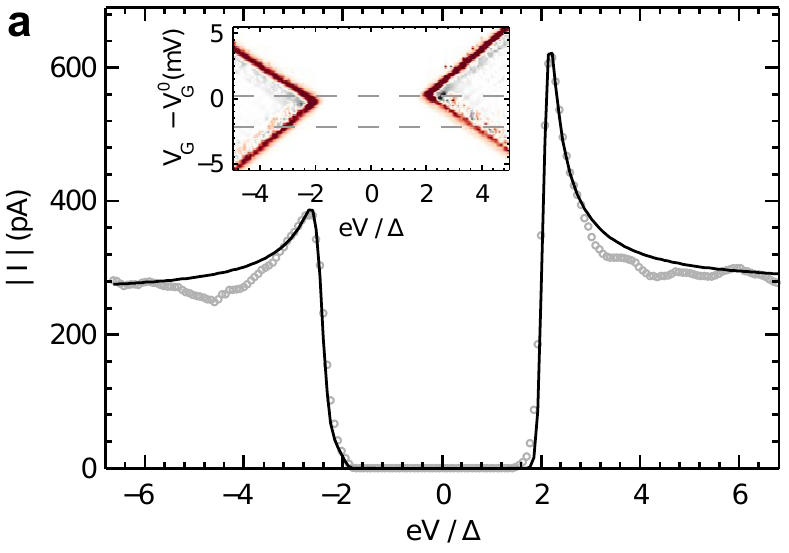}
		\includegraphics[width=0.9\columnwidth]{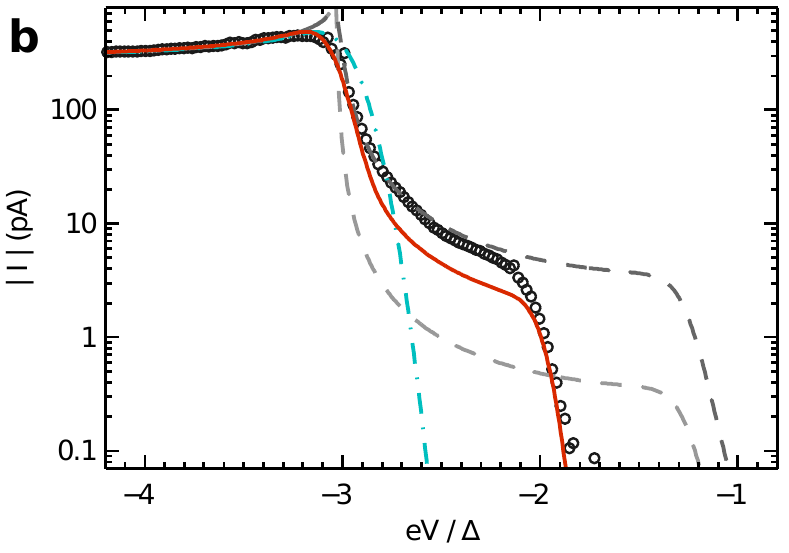}
	\caption{(a) $|I(V)|$ trace for $V_G\approx V_G^0$ in the more symmetric device S. The line is a fit using the model developed in the text. Note the asymmetry of both the on-state currents and the shape of the coherence peak replicas at opposite voltages, which are well captured by the model. Inset: $\partial I / \partial V$ map of device S near $V_G^0$. Differential conductances are shown red when large and positive, white for close to zero and black for negative. The dashed lines indicate the gate voltages at which the $I(V)$ traces displayed in this figure (a\& b) are taken.  (b) Logarithmic scale representation of $|I(V)|$ trace at $V_G-V_G^0=-2.17$ mV at the conductance threshold (symbols). Best fits assuming cotunneling model (red line), sequential tunneling with a 10 $\mu$V gaussian broadening accounting for noise (cyan dashed dotted line) and inclusion of a Dynes parameter ($\eta=10^{-2}\Delta$, dark grey dashed line; $\eta=10^{-3}\Delta$, light grey dashed line) are also shown.}
	\label{broadening}
\end{figure}

The two samples that we report on here are characterized in Table 1. Both samples have charging energies $U>100$ meV and display the well known transport properties of weakly coupled S-QD-S transistors \cite{Ralph95}. Following the orthodox single electron transistor picture \cite{Beenakker91,Averin91,Bonet02} the device is characterized by two tunnel couplings $\gamma_{i}$ to the source and drain leads  and three capacitances ($C_i$) to the gate ($G$), source ($S$) and drain ($D$) electrodes respectively, see Fig. \ref{device}. The dimensionless gate coupling parameter $\alpha=C_G/C_{\Sigma}$ (with $C_{\Sigma}=C_G + C_S + C_D$) indicates the efficiency of the conversion of the gate voltage applied into variations of the QD chemical potential; values of $\alpha$ in the $10^{-2}$ range are usual in electromigration junctions.
Because of the extremely high charging energy however, only one charge degeneracy point is accessible in each device, at $V_G^0$. We name $N$ and $N+1$ the number of electrons on the QD at $V_G$ smaller/larger than $V_G^{0}$ respectively.
All devices we have measured showed an extra gate independent contribution to the current. We attribute this to a direct superconductor-to-superconductor (SIS) transport channel, shunting the QD due to the narrowness of the gap produced by electromigration. This small contribution corresponding to a shunt resistance $R_s$ in the G$\Omega$ range can be well fitted as a standard SIS tunnel junction and subtracted from the data. At the degeneracy point, a spectroscopic gap of total width $4\Delta \approx1$ meV persists in the $I(V)$ traces (Fig. \ref{device}c and \ref{broadening}a). This is because elastic single quasi-particle transport at low temperatures through a S-QD-S device requires {\it a minima} the filled states of one lead to be above  the empty states of the opposite lead, that is $ |V |>2\Delta/e$. When a magnetic field is applied, the spectroscopic gap gradually decreases and disappears above 500 mT (Fig. \ref{device}d). 

Discrete and sharp peaks in the $\partial I / \partial V$ maps demonstrate that individual single orbital levels are addressed when sweeping the bias (Fig. \ref{device}c). The QD level spacing $\delta E$ exceeds 1 meV, in agreement with earlier works on similar devices with normal leads \cite{Kuemmeth08}. At sufficiently high magnetic fields, the Zeeman splitting of the QD orbital level involved, proportional to the applied magnetic field, is well resolved in the conductance data (Fig. \ref{device}d). Because of the strong asymmetry of the tunnel rates to both leads of device A, shown in Fig. \ref{device}, this splitting is only seen along a single edge of the $N$ occupied Coulomb diamond. Since the first excited state of the $N \rightarrow N+1$ transition does show Zeeman splitting, we can conclude that $N+1$ is odd in this device \cite{Ralph95}. The associated electron Land\'e factor is 2.4$\pm0.3$. This slightly larger value than previously reported results for gold nanoparticle junctions \cite{Kuemmeth08} may be due to the fact that the Zeeman energy is determined here at rather small fields.

\begin{table}[t]
	\centering
	\begin{tabular}{ l | c c c c c c c c}
		 &  $R_{s}$ & $\alpha$ & $\,C_D/C_S\,\,$ & $I_{+}\,\,$ & $|I_{-}|$ & $\hbar \gamma_S$ & $\hbar \gamma_D$  \\ 
		  &  $(\mathrm{G\Omega})$ & & & (pA) & (pA) & ($\mu$eV) & ($\mu$eV) \\
		\hline
		A &  0.5 & 0.05 & 0.70 & 166 & 92 & 5.2 & 0.4  \\
		S &  1 & 0.09 & 1.07 & 290 & 250 & 2.1 & 1.4 \\
	\end{tabular}
	\caption{Device parameters of devices A and S. $R_{s}$ is the shunt resistance, $\alpha$ the gate coupling and $C_D/C_S$ the capacitive asymmetry.  For the definition of the on-state currents ($I_{+}$  and $I_- $) and the tunnel couplings $\gamma_{S,D}$, see text.}
	\label{parameters}
\end{table}

In the normal state of the leads, the on-state currents of the QD device $I_{\pm}$ (where $\pm$ reflects the sign of the applied bias voltage) are directly related to the tunnel couplings $\gamma_{S, D}$ \cite{Bonet02}.
The same is true with superconducting leads at bias voltages well above $2\Delta /e$. 
Notably, both devices presented here
have comparable on-state currents, although quite different coupling asymmetry (Table 1): sample A has quite asymmetric tunnel couplings, as opposed to the more symmetric sample S. Further, the tunnel couplings, in the $\mu$eV range, are slightly smaller than $k_B T\approx 10\,\, \mu$eV. They are also significantly below all other energy scales, that is, $\Delta\approx 260$ $\mu$eV, which is itself much less than the level spacing $\delta E$ (several meV) and the charging energy $U>50$ meV. This very well separated hierarchy of energy scales is a rather unique situation in that it allows for a truly quantitative analysis of charge transport through a single quantum level.

\begin{figure}[t]
\vspace{-0.9cm}
	\hspace{-1.3cm}
	\includegraphics[width=1.1\columnwidth]{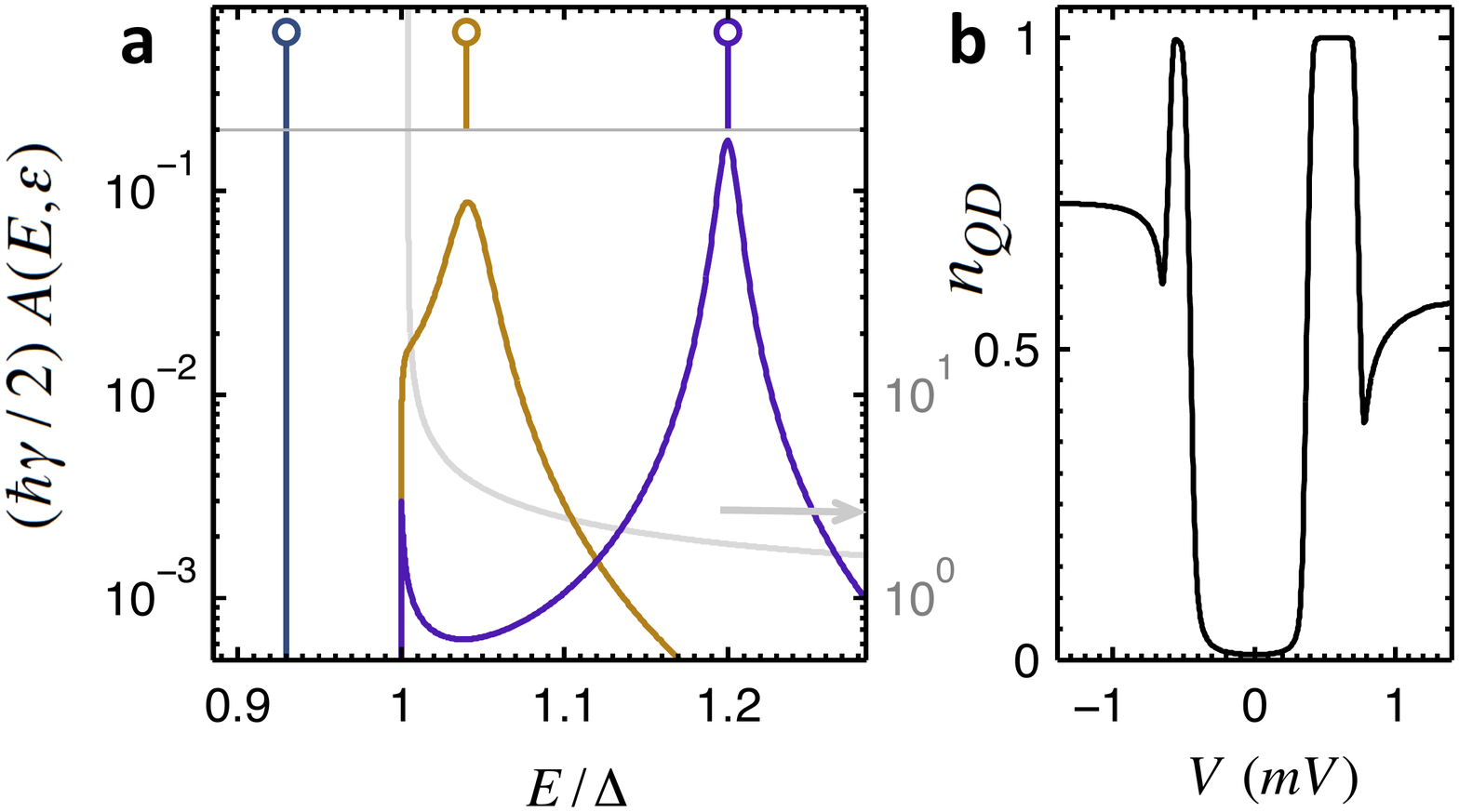}
		\vspace{-1.3cm}
	\caption{ (a) Calculation of the normalized spectral function of a single quantum level tunnel coupled (with $\hbar \gamma=8\cdot 10^{-3}\Delta$) to a superconducting reservoir for three different bare level energies $\epsilon$ (indicated by the position of the circles on the top part). The grey line shows $\rho(E)$ as a reference. (b) Occupation $n_{QD}$ of the S-QD-S device as a function of $V$ for $\alpha (V_G-V_G^0 )= -65 \mathrm{\mu eV}$, as calculated from Eq. (\ref{filling-b}). We assume $\Delta=260$ $\mu$eV and tunnel couplings as for device S (Table \ref{parameters}).}
	\label{calculations}

\end{figure}

When coupled to a shapeless continuum, a discrete quantum level  decomposes into a lorentzian spectral function. In the presence of a non-trivial (normalized) superconducting density of states $\rho(E)$ in the leads, given by the real part of $| E|/ \sqrt{E^2-\Delta^2}$, hybridization can be described via an effective tunnel coupling parameter. Consequently, the tunnel coupling itself depends on the relative position of  $\epsilon$ with respect to the lead's chemical potential, namely  ${\tilde \gamma}(\epsilon)=\hbar \gamma\,  \rho(\epsilon)/2$. A Green's function based calculation yields the hybridized QD spectral function $A(E,\epsilon)=(1/\pi)\,{\cal I}m \left[ (E-\epsilon-\Sigma)^{-1} \right]$, which is shown for various bare QD energy levels $\epsilon$ in  Fig. \ref{calculations}a. Here ${\cal I}m$ represents the imaginary part and $\Sigma= -\hbar \gamma |E| / \sqrt{\Delta^2-E^2}$ is the self-energy. Marked distortions from a lorentzian line shape (of width $\hbar \gamma$) are manifest as $\epsilon$ approaches the superconducting gap edge. At $E = \Delta$, the spectral density falls abruptly to zero; bare QD levels below the gap edge are virtually unaffected by the tunnel coupling.

This approach allows to describe cotunneling contributions to charge current which, by truncating higher-order Green's functions and assuming a single spin-degenerate level with $U\rightarrow +\infty$, is found to be\cite{Kang98}

\begin{equation}
\begin{split}
I(V) & =  \frac{4e}{h}\, (2-n_{QD}) \int_{-\infty}^{+\infty}  \left[ n_S(E) -n_D(E+eV)   \right] \\
 & \\
 & \times  \frac{{\tilde \gamma}_S(E)\, {\tilde \gamma}_D(E+eV)}{(E-\epsilon)^2+\left[ {\tilde \gamma}_S(E)+{\tilde \gamma}_D(E+eV)  \right]^2}\, \, dE.\\
\end{split}
\label{kang}
\end{equation}
Here the $n_{i}$ are the respective filling factors of the states in the source, drain and quantum dot. 
In the sequential tunneling limit ($\gamma_i \to 0$), the lorentzian functional inside the integrand reduces to a Dirac peak and $|I(V)|$ replicates the leads' superconducting DOS \cite{Ralph95}. Note that we have not chosen the usual symmetric biasing convention ($V_S=-V_D=V/2$) but we write $V_S=V$ and $V_D=0$. This convention has the advantage of being closer to the experimental reality. We take into account the explicit capacitive dependence of de QD energy $\epsilon$ on $V$, that is $\epsilon(V,V_G)=\alpha(V_G-V_G^0)+V C_S/C_{\Sigma}$.

The prefactor $(2-n_{QD})$ plays an essential role in that it accounts for Coulomb interactions on the quantum dot \cite{Kang98}. A similar expression to Eq. (\ref{kang}) without this prefactor is found when on-dot interactions are neglected \cite{Yeyati97}. 
A general expression of the QD filling factor $n_{QD}$, accounting for cotunneling  and valid for arbitrary bias voltage and leads DOS, is given by \cite{Kang98b} 

\begin{equation}
n_{QD}= 2 \left( \pi + \int_{-\infty}^{+\infty}   \frac{{\tilde \gamma}_S\, n_S - {\tilde \gamma}_D\, n_D }{(E-\epsilon)^2+\left( {\tilde \gamma}_S+{\tilde \gamma}_D  \right)^2}\, dE \right)^{-1},
\label{filling-b}
\end{equation}
where the functions $\tilde \gamma_i$ and $n_i$ in the integrand are evaluated at $E$ and $E+eV$ for $i=S,D$ respectively. The spin-degeneracy of the QD level is summed out in the above expression. The strong energy dependence of $n_{QD}$ (Fig. \ref{calculations}b) is associated to the leads' DOS.
In the on-state (at high bias) the expression of $n_{QD}$ reduces to 

\begin{equation}
n_{QD}^{on}=2\frac{\gamma_D\,n_D+\gamma_S\,n_S}{(1+n_D) \gamma_D+(1+n_S)\gamma_S},
\label{filling}
\end{equation}
as given by rate equations \cite{Bonet02}. Note that in symmetric tunnel coupling conditions ($\gamma_S=\gamma_D=\gamma$) this expression yields $n_{QD}^{on}=2/3$ and, using Eq. (\ref{kang}), $I_{\pm}=\pm (2/3)e \gamma$. 

The current-voltage characteristics of device S near the $(N,N+1)$ degeneracy point are shown in Fig. \ref{broadening}a. As expected, $| I(V) |$ is a smeared replica of the leads superconducting density of states. The small asymmetry in the capacitive coupling to the leads (see inset Fig. \ref{broadening}a) is strongly reflected in amplitude and position of the coherence peaks' replicas. The cotunneling model, leading to Eqs. (\ref{kang}) and (\ref{filling-b}), provides a very good quantitative description of the data. In particular, the model captures well the drain-source asymmetries in terms of $I_{\pm}$ and broadening at the current onset.

Clear evidence of spectral hybridization is found in the behavior of the sub-threshold current measured off charge degeneracy.
At finite gate detuning, there is a non-negligible contribution to current below the threshold bias voltage, which is highlighted on a logarithmic scale in Fig. \ref{broadening}b. The cotunneling model describes well the long tail of the subthreshold current, as well as its sudden rapid drop for $|V_b|<2\Delta/e$. The slight theoretical underestimation of the subthreshold current can most probably be assigned to non-equilibrium effects, both in the leads' and in the QD filling factors \cite{Kang15}, that are neglected here. The overall shape of the subthreshold cotunneling contribution combined with the current far above threshold provide a self-consistent estimate of the tunnel couplings to both leads (Table 1), which are the only free parameters. The steep slope of the $| I(V) |$ curves at  $V=\pm 2\Delta / e$ is independent on the dot level and tunnel coupling and reflects the experimental bias noise of $\sim 10\, \mu$V.   Notably, the effect of temperature on the above description only appears with weighting $\sim \exp (-min(\Delta,\delta E)/k_BT)$ and thermal smearing is thus completely negligible.

Alternative explanations of the subthreshold current behavior could be thought of in terms of noise or an effective broadening of the leads' superconducting DOS. A sequential-tunneling calculation neglecting hybridization and assuming 10 $\mu$V gaussian noise is missing the slow lorentzian decay of the current (Fig. \ref{broadening}b). While the intrinsic broadening of the aluminum DOS  is known to be extremely small ($<10^{-6}\, \Delta$) \cite{Pekola10}, we have alternatively tried fitting the experimental data assuming a Dynes-type correction to the superconducting aluminum density of states \cite{Dynes78} in which an imaginary term $i\eta$ is added to $E$ in the definition of $\rho(E)$. 
The assumption of an unrealistically large $\eta \sim \gamma$ can to some extent account for the subthreshold decay of the current (Dynes corrections also appear as a lorentzian contribution in this regime). A Dynes-type superconducting DOS is however unable to account for neither the shape of the coherence peaks nor the rapid fall-off at $V=\pm 2\Delta /e$ (Fig. \ref{broadening}b). This confirms that spectral hybridization associated to cotunneling is the dominant mechanism at work, accounting for all aspects of spectroscopic broadening in our data.

In conclusion, we have shown that while spectroscopy with superconducting probes is known to help avoiding thermally induced smearing, the main source of spectroscopic broadening is then associated to an enhanced invasiveness of the tunnel coupling, producing hybridization. The tunnel coupling to the leads has thus to be kept very small for well-resolved spectroscopies, implying small tunnel currents. This is of particular relevance to a panel of cryogenic STM experiments on low dimensional nano structures, in which superconducting tips can provide sub-$k_BT$ spectroscopic resolution \cite{Franke11,Heinrich13,Nadj14}.

This work was funded by the EC in the frame of the FET-open project INFERNOS. Samples were fabricated at the Nanofab facility at Institut N\'eel - CNRS. We thank D. Basko, K. Kang, J. Pekola, F. Hekking, B. Sac\'ep\'e and F. Balestro for help and stimulating discussions.

\bibliography{./SQDS-main}

\end{document}